\documentclass{aa}
\usepackage{psfig}

\newcommand{\nh}{N_{\rm H}}
\newcommand{\cmsq}{{\rm atoms\,cm}^{-2}}
\newcommand{\ergs}{{\rm erg\,s}^{-1}}

\newcommand{\ergcms}{{\rm erg\,cm}^{-2}{\rm s}^{-1}}
\newcommand{\ltap}{\mathrel{\hbox{\rlap{\lower.55ex \hbox {$\sim$}}
                   \kern-.3em \raise.4ex \hbox{$<$}}}}
\newcommand{\gtap}{\mathrel{\hbox{\rlap{\lower.55ex \hbox {$\sim$}}
                   \kern-.3em \raise.4ex \hbox{$>$}}}}

\begin{document}

\title{A four-hours long burst from Serpens\,X$-$1}

\author{R.\ Cornelisse\inst{1,2} \and  E. Kuulkers\inst{1,2} \and 
        J.J.M.\ in 't Zand\inst{2,1} \and  F.\ Verbunt\inst{2}
	\and J.\ Heise\inst{1}}
\offprints{R.\ Cornelisse}
\mail{R.Cornelisse@sron.nl}

\institute{SRON National Institute for Space Research, Sorbonnelaan 2, 
              3584 CA Utrecht, The Netherlands  
         \and  Astronomical Institute, Utrecht University,
              P.O.Box 80000, 3508 TA Utrecht, The Netherlands 
                }

\date{\today / Accepted date}

\abstract{During a serendipitous observation of the
BeppoSAX Wide Field Cameras, a very long Type\,I X-ray
burst was observed from the low mass X-ray binary Serpens 
X$-$1. The burst lasted for approximately 4 hours and had an
exponential decay time of $69\pm2$ min (2-28 keV). The 
bolometric peak-luminosity is ($1.6\pm0.2)\times10^{38}$ $\ergs$ 
and the fluence ($7.3\pm1.4)\times10^{41}$ erg. The first 
'normal' Type\,I burst was observed 34 days after  
the superburst. This is in rough agreement with recent predictions
for unstable carbon burning in a heavy element ocean.
}

\maketitle

\section{Introduction}

Since the first report of a very long thermo-nuclear
X-ray burst in 4U\,1735$-$44 (Cornelisse et al. 2000), 
six more of these so-called 'superbursts' have been noted 
(Strohmayer 2000;  Heise et al. 2000; Wijnands 2001; Kuulkers 2001).
The superbursts have the following common properties:  
a long duration of a few hours, a large burst energy 
($\sim10^{42}$ erg) and a persistent pre-burst luminosity 
between 0.1 an 0.3 times the Eddington limit $L_{\rm Edd}$ 
(Wijnands 2001). In addition, all superbursts are known 
Type\,I X-ray bursters.

Apart from its duration, a superburst shows all the 
characteristics of a Type\,I X-ray burst, namely:  
the lightcurve has a fast rise and exponential decay; 
spectral softening occurs during the decay; black-body 
radiation describes the burst X-ray spectrum best. Normal 
Type\,I bursts can be explained very well by unstable He 
and/or H fusion on a neutron star surface (for reviews 
see e.g. Lewin et al. 1993, 1995; Bildsten 1998). In contrast, 
the superbursts are possibly due to unstable carbon fusion in 
layers at larger depths than where a typical Type\,I burst 
occurs (Cumming \& Bildsten 2001; Strohmayer \& Brown 2001). 

In this paper we report the detection of one of the seven
superbursts, namely from
the X-ray source Serpens\,X$-$1 (Ser\,X$-$1), as observed with 
one of the Wide Field Cameras (WFC) onboard BeppoSAX. Ser\,X$-$1 
is a relatively bright persistent X-ray source discovered in
1965 (Friedmann et al. 1967). Over 100 'normal' Type\,I 
bursts have been reported from Ser\,X$-$1 (e.g. Swank 
et al. 1975, Sztajno et al. 1983, Baluci\'nska 1985). The
proposed optical counterpart is MM Ser (Thorstensen et 
al. 1980). Wachter (1997) showed that the object is a 
superposition of two stars, and that no clear period 
could be derived from a photometric study. A distance 
of 8.4 kpc derived from Type\,I bursts is given by Christian \&
Swank (1997).

In this paper we describe the observation and properties 
of the Ser\,X$-$1 superburst, and discuss this in context
to the other superbursts reported so far. (The occurrence
of this burst was first mentioned in Heise et al. 2000.)

\section{Observations}

The Wide Field Cameras are two identical coded mask cameras
onboard the Italian-Dutch satellite BeppoSAX (Jager et al.
1997, Boella et al. 1997). An overview
of the characteristics of the WFC is given in Jager et al. (1997).

Most WFC observations are done in secondary mode. These are 
arbitrary sky-pointings except that they are perpendicular
to the direction of the target to which the Narrow Field Instruments 
onboard BeppoSAX are pointed, and dictated by solar constraints. 
During the first half of 1997, the WFC observed Ser\,X$-$1 for a 
total of 411 ks (corrected for earth occultation and south Atlantic 
anomaly passages), distributed over 12 observations. In Table\,
\ref{observation} an overview of all these observations is 
given. 

During this period, there were also two RXTE Proportional
Counter Array (RXTE/PCA) observations. The RXTE/PCA is an 
array of 5 co-aligned Proportional Counter Units (PCU). In 
Jahoda et al. (1996) a detailed description is given of the 
instrument. All PCU's were on during the observations. We 
use standard 1 data for our analysis. Also on-board RXTE
are three Scanning Shadow Cameras with a $6^\circ\times90^\circ$
field of view forming the All-Sky Monitor (ASM; Levine 1996).
We use the data products provided by the RXTE/ASM
team at the MIT web-pages.

\begin{table}[h]
\caption{Overview of the WFC and PCA observations of Ser\,X$-$1 
between February 23 and May 13 1997. The exposure time is 
corrected for earth occultation, South Atlantic Anomaly passages 
and other data gaps. The WFC observation of Ser\,X$-$1 prior to this
period was on November 6 1996 (MJD 50393) and the one following
on August 22 1997 (MJD 50682). 
\label{observation}}
\begin{tabular}{lccc}
\hline
& start (MJD) & end (MJD) & net exposure (s)\\
\hline
WFC & 50502.29 & 50503.37 & 43437\\
    & 50506.82 & 50508.02 & 32237\\
    & 50508.61 & 50509.29 & 31180\\
    & 50513.16 & 50513.86 & 29020\\
    & 50517.70 & 50518.00 & 15306\\
    & 50518.26 & 50519.36 & 33699\\
    & 50529.63 & 50530.20 & 15496\\
    & 50541.26 & 50542.08 & 36302\\
    & 50542.63 & 50543.37 & 33310\\
    & 50554.62 & 50555.77 & 60707\\
    & 50563.69 & 50565.86 & 75693\\
    & 50581.47 & 50581.63 & 4652\\
PCA & 50535.5 & 50539.9 & 34560\\
    & 50554.4 & 50554.7 & 25920\\
\hline
\end{tabular}
\end{table}

\section{Data analysis \& Results}

On February 28, 1997 a flare-like event was observed which lasted
for almost 4 hours. In Fig.\,\ref{lightcurve} we show the RXTE/ASM
lightcurve
of Ser\,X$-$1 over a period of 5 years (a), and an expanded lightcurve 
during spring 1997 (b). The flare was 
observed after BeppoSAX came out of earth occultation on MJD 
50507.075.  In Fig.\,\ref{lightcurve}c and d a detailed view of the
flare is shown. The rise to maximum was missed. After the satellite 
came out of earth occultation 
a flat top is observed before the decay starts, so it appears that 
the peak is covered. The flare shows an exponential decay. In 
Fig.\,\ref{lightcurve}e
we see that spectral softening occurs during the decay. This is 
indicative for Type\,I bursts. In Table\,\ref{results} we summarize 
the exponential decay times from fits in different energy bands. 

We divided the observation in which we discovered the superburst
(see Fig.\,\ref{lightcurve}c and d) in three different intervals 
before, during and after the superburst, as indicated by dashed 
lines in Fig.\,\ref{lightcurve}c. In the first and last interval 
we fit the persistent flux with solely an absorbed bremsstrahlung 
spectrum. During the superburst we used a sum of the (persistent) 
bremsstrahlung and black-body radiation to describe the flux, taking 
the persistent bremsstrahlung emission as fixed at the average 
level of the spectral fits before and after the burst. We tried several 
other spectral models for the persistent emission, like a cut-off power-law 
and a disk black-body. The derived fluxes  did not change significantly 
for the different models, and the best fit during the 
superburst is given by the bremsstrahlung model. In all our 
fits we fixed the absorption column at $\nh=0.5\times10^{22}$ $\cmsq$ 
(Christian \& Swank 1997). We also added a black-body component
to model the emission after the burst. However, an F-test showed 
that this extra component did not improve the fit significantly
($\simeq2.5\sigma$ probability).

\begin{figure}[ht] 
\psfig{figure=lightcurve_ser.epsi,height=14.5cm}
\caption*{{\bf a)} RXTE/ASM lightcurve of Ser\,X$-$1.
Each bin is a 2 day average, and normalized to that of the Crab; 
only data with more than one dwell per day were included. The 
time span between the dashed lines has been blown up in {\bf b)},
where we also show the data of the WFC (2-28 keV), and RXTE/PCA 
(indicated by small horizontal bars underneath). The WFC bins are one orbit 
averages and the PCA bins are 5 min averages; data points 
are connected to guide the eye when less than 0.2 days apart. The ASM 
points are one day averages (with more than one dwell per day). 
All observations are normalized to the Crab. A sharp increase can 
be seen at MJD 50507. The vertical tick-marks at the middle-right
indicate the occurrence of 'normal' Type\,I bursts. Panels {\bf c-e}
show again a blow-up, indicated by the dashed line in {\bf b)}, 
of the superburst in two different energy-bands, i.e. 2-5 keV {\bf (c)}
and 5-28 keV {\bf (d)}, and their ratio in {\bf (e)}.
Each bin is 5 min. In panel {\bf c)} individual ASM dwells are also 
over-plotted.
\label{lightcurve}}
\end{figure}

\begin{figure}[h] 
\psfig{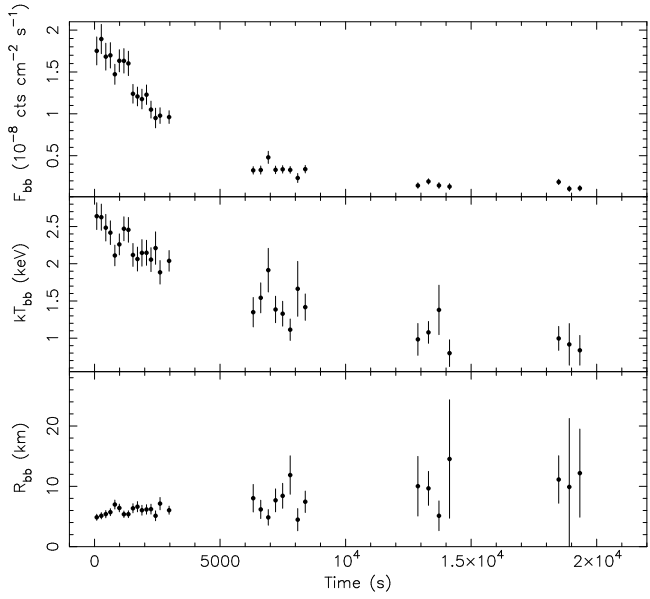}
\caption{Results from time resolved spectroscopy from 
the beginning of the burst back to the persistent emission 
level. We show the black-body flux ($F_{\rm bb}$), black-body
temperature ($kT_{\rm bb}$) and black-body radius ($R_{\rm bb}$)
of the burst, respectively.
We have taken the persistent emission level, modeled 
by a bremsstrahlung spectrum, as fixed at the same values 
as used with the spectral fit of the total burst (see Table\, 
\ref{results}). Also the absorption column is fixed at $\nh=
0.5\times 10^{22}$ $\cmsq$ (Christian \& Swank 1997). During 
the first part of the burst each bin is 180 s. After the first 
gap due to earth occultation the bin size becomes 300 s and the 
last two intervals have bin sizes of 420 s. The black body radius 
is for a source at 8.4 kpc distance. \label{spectrum}}
\end{figure}

In Fig.\,\ref{spectrum} the results of the time-resolved 
spectral fits are shown. It is seen that both the black-body 
flux and the temperature drop exponentially, while the radius 
stays nearly constant. This is typical for normal Type\,I bursts.

An increase in the RXTE/ASM count-rate
is observed on MJD 52120.15. This increase lasted for several dwells, 
having a total duration of at least 8.6 minutes and less then 2.3 hrs.
Spectral softening can not be proven. This flare-like event may be 
another superburst.

\begin{table}[h]
\caption{Fit results of the superburst. The top panel 
shows the decay times for exponential decay fits in  
different passbands. The next panels shows the spectral 
fit results for the persistent emission before, after
and during the burst, respectively. 
A bremsstrahlung model is employed to describe the persistent 
emission. The burst-emission is described by a sum of
the bremsstrahlung and black-body spectrum, taking the
bremsstrahlung emission fixed at the average of the 
spectral fits before and after the burst.
For all spectral fits we fixed the absorption column 
at the value of $\nh=0.5\times10^{22}$ $\cmsq$ (from 
Christian \& Swank (1997)). 
\label{results}}
\begin{tabular}{lc}

\hline
{\bf exponential decay}\\
$\tau_{2-28 \rm\, keV}$ (min) $(\chi^2_\nu$, d.o.f.) & 69$\pm$3 (1.2, 53) \\
$\tau_{2-5 \rm\, keV}$  (min) $(\chi^2_\nu$, d.o.f.) & 108$\pm$12 (1.2, 51)\\
$\tau_{5-28 \rm\, keV}$ (min) $(\chi^2_\nu$, d.o.f.) & 52$\pm$2 (1.4, 56) \\
\hline
{\bf brems before}\\
$kT_{\rm br}$ (keV)               & 7.3$\pm$0.4\\
$F_{2-28 \rm\, keV}$    ($10^{-9} \ergcms$) & 5.0$\pm$0.2\\
$F_{\rm bol}$ ($10^{-9} \ergcms$) & 8.3$\pm$0.2\\
$\chi_\nu^2$  (d.o.f.)            & 0.8 (27)\\
\hline
{\bf brems after}\\
$kT_{\rm br}$(keV)                & 6.2$\pm$0.4\\
$F_{2-28 \rm\, keV}$    ($10^{-9} \ergcms$) & 5.7$\pm$0.3\\
$F_{\rm bol}$ ($10^{-9} \ergcms$) & 10.1$\pm$0.4\\
$\chi^2_\nu$  (d.o.f.)            & 1.0 (27)\\
\hline
{\bf brems \& black-body during}\\
$kT_{\rm bb}$  (keV)               & 2.4$\pm$0.1\\
$R$     (km at 8.4 kpc distance)    & 3.6$\pm$0.3\\
$F_{2-28 \rm\, keV}$     ($10^{-9} \ergcms$) & 6.1$\pm$1.1\\
$F_{\rm bol}$  ($10^{-9} \ergcms$) & 6.3$\pm$1.1\\
$L_{\rm peak}$ ($10^{38} \ergs$) & 1.6$\pm$0.2\\
$E_{\rm b}$ ($10^{41}$ erg)        & 7.8$\pm$1.4\\
$\chi^2_\nu$   (d.o.f.)            & 1.7 (26)\\
\hline
\end{tabular}
\end{table}

\section{Discussion}

The flare-like event from Ser\,X$-$1 shows, except for
its duration, all characteristics of a Type\,I burst.
There are now six sources showing superbursts lasting 
for hours up to half a day. This superburst
occurred at a persistent flux level of 0.21$L_{\rm Edd}$, 
for $L_{\rm Edd}=2\times10^{38}$ $\ergs$. This is comparable 
to that for the other superbursts. The fact
that the persistent luminosity before the superburst
is similar in the 6 sources may be an 
observational selection effect. As suggested by Cumming 
\& Bildsten (2001), superbursts can occur at higher luminosities,
where they are more difficult to detect due to the smaller
contrast with the persistent flux. The total energy emitted 
during the burst, $E_{\rm b}=(7.3\pm1.4)\times10^{41}$ 
erg, and the exponential decay time, are also 
comparable to those of the other superbursts. 

Cornelisse et al. (2000) noticed the absence of normal Type\,I 
bursts in the regular burster 4U\,1735$-$44 throughout the 
9-day observation. In the case of KS\,1731$-$260, no Type\,I 
bursts were seen after the superburst, whereas they were present 
beforehand (Kuulkers et al. 2001). From
Fig.\,\ref{lightcurve} we see a similar effect in 
Ser\,X$-$1. The first normal burst is observed about 34 days after 
the superburst, at MJD 50541.32. After that, Ser\,X$-$1 is seen to 
burst rather regularly, having a total of seven bursts in 210.7 ks
(about 3 bursts per day). For this burst-rate the Poisson distribution
predicts a probability 
of 24\%\ that no bursts are detected in the 55 ks WFC observations 
before the superburst, so it is possible that the lack of observed
bursts before the superburst is due to chance. The probability 
that the absence of ordinary bursts in the 34 days after the superburst 
is due to chance is 1.2\%. 

We also performed Monte Carlo simulations to verify in more
detail the significance of finding no bursts around the superburst.
We randomly varied the burst waiting time in an interval symmetric
around 0.38 day (i.e. the average waiting time), and taking as the 
lower limit 0.04 day (the lowest waiting time observed by Sztajno et al. 
1983). We then determined
the number of bursts in: 1) the observing window before the
superburst, and 2) the observing windows between the superburst 
and the occurrence of the first normal Type\,I burst. By doing 
$10^6$ simulations for each window we found that the expected 
number of bursts for window 1) is $1.5\pm1.1$ and the chance of 
observing no bursts is 14\%. For window 2) the expected number 
of bursts is $4.9\pm1.9$ and the chance of observing no bursts 
is only 0.4\%. We conclude that the absence of bursts after the
superburst is significant.

Fig.\,\ref{lightcurve} shows that after the superburst 
the persistent flux-level in the low energy passband is higher 
then before the burst, while in the high energy passband no 
significant increase is observed. This is also visible in the
burst from 4U\,1735$-$44 (Cornelisse et al. 2000), but less obvious 
in KS\,1731$-$260 (Kuulkers et al. 2001). This offset
could be due to the heating of the neutron star atmosphere after the
superburst, and could be present in the spectrum as a black-body 
component. However, for both Ser\,X$-$1 and KS\,1731$-$260 no significant 
black-body contribution can be proven. We re-analyzed the off-set 
emission after the superburst of 4U\,1735$-$44 (interval 
MJD 50318.55-50319.0). Here, $\chi^2=38.2$ 
(27 d.o.f.) for the absorbed bremsstrahlung model, which improves to 
$\chi^2=17.2$ (25 d.o.f.) when a black-body component was added. 
Performing an F-test shows that this is a significant ($\simeq4\sigma$
probability) improvement. This may indicate that, at least for
4U\,1735$-$44, the neutron star 
atmosphere is still cooling $\simeq0.5$ day after the superburst.

Cumming \& Bildsten (2001) propose that these bursts 
are due to unstable carbon burning in an ocean of heavy elements, e.g. 
$^{104}$Ru. This can explain the observed durations and 
fluences for these superbursts. Also, the recurrence time is 
roughly consistent with that found for 4U\,1636$-$53 (Wijnands 2001). 
A waiting time is predicted before the normal
Type\,I bursts start again; $t_{\rm wait}\simeq5t_{\rm cool}$,
where the cooling time is a function of mass accretion rate (Cumming 
\& Bildsten 2001). In the case of Ser\,X$-$1, for which 
the persistent flux is roughly 
0.2$L_{\rm Edd}$, a waiting time of $\sim$15 
days is predicted. Given the rough estimates above, and the small
observational coverage between 15 days and 34 days after the 
burst, the predicted waiting time is consistent with the observation.

If the increase in the RXTE/ASM data is due to a superburst, an
upper limit of 4.4 years on the recurrence time of superbursts
in Ser\,X$-$1 can be estimated. This timescale is comparable 
to the 4.7 years found for 4U\,1636$-$53 (Wijnands 2001).

\begin{acknowledgements}
We thank Lars Bildsten for critically reading the manuscript, and 
Rudy Wijnands for pointing out the existence of the RXTE/ASM flare. 
The BeppoSAX satellite is a joint Italian and Dutch program. We made
use of the quick-look results provided by the RXTE/ASM team.
\end{acknowledgements}

\end{document}